# Electric Polarization and Magnetic Properties of $(NH_4)_{1-x}K_xI$ (x = 0.05–0.17)

Yi Yang Xu,[a,#] Lei Meng,[a,#] Miao Miao Zhao,[a,#] Chu Xin Peng,[a] and Fei Yen[*,a]

[a]State Key Laboratory on Tunable Laser Technology, Ministry of Industry and Information Technology Key Laboratory of Micro-Nano Optoelectronic Information System and the School of Science, Harbin Institute of Technology, Shenzhen, University Town, Shenzhen, Guangdong 518055, P. R. China

**Abstract:** While all of the polymorphs of pure $NH_4I$ and $KI$ are non-polar, we identify that $(NH_4)_{0.95}K_{0.05}I$ is ferroelectric and $(NH_4)_{0.87}K_{0.13}I$ and $(NH_4)_{0.83}K_{0.17}I$ are pyroelectric through measurements of their pyroelectric current and complex dielectric constant. The order to disorder phase transitions occur near 245 K. Magnetic susceptibility measurements indicate that the proton orbitals of the $NH_4^+$ continue to become ordered in the ground state in the $(NH_4)_{1-x}K_xI$ system up to $x \leq 0.17$. The polar phases are proposed to stem from $K^+$ ions disrupting the symmetry of proton-orbital-lattice interactions between the $NH_4^+$ and $I^-$ ions. Our work introduces a new pathway for the ordered phases of ammonium-based compounds to potentially become ferroelectric.

[#] These authors contributed equally.
[*] Corresponding author: Fei Yen, fyen@hit.edu.cn, Tel: +86-1343-058-9183

## 1. Introduction

The order-disorder transition temperature $T_{\beta-\gamma}$ = 232 K of $NH_4^+$ ions in ammonium iodide $NH_4I$ was recently identified to possibly originate from orbital magnetic resonant interactions between intermolecular protons ($H^+$) [1]. As shown in Fig. 1a, the $NH_4^+$ cations are ~7 Å apart from each other and reorient at rates of ~$10^{-12}$ s [2–6]. Reorientations of the $NH_4^+$ involve protons orbiting in concert around a central stationary nitrogen atom. Due to the periodic nature of the potential barriers the $NH_4^+$ only exhibit two energetically favorable types of reorientations [6]: a) by 120° ($C_3$) clockwise or anticlockwise along each of the four N–H bonds (Fig. 1b); and b) by 180° ($C_2$) or 90° ($C_4$) along the *a*, *b*, *c* axes of the lattice (Fig. 1c). Each type of reorientation has its own orbital magnetic moment $\mu_p$ (perpendicular to the plane of the proton orbital) since rotating molecules generate a magnetic field [7,8]: the electrons reside near the nitrogen atom and the protons at the edges trace out current loops. In the $\beta$ and $\gamma$ phases, which are cubic of the CsCl type and slightly tetragonal, respectively, each $NH_4^+$ tetrahedron is surrounded by eight nearly-equidistant $I^-$ so that the equilibrium positions of the protons lie more or less midway between nitrogen and iodine atoms [9]. This allows for each $NH_4^+$ to take 1 out of 2 possible types of spatial configurations, *A* and *B*; a $C_4$ reorientation switches *A* to *B* or vice versa while $C_3$ and $C_2$ reorientations allow the $NH_4^+$ to retain its original configuration (Fig. 1a). In the low temperature $\gamma$ phase, the $NH_4^+$ "order" in an *ABAB*… anti-parallel fashion along the *ab* planes [9]. We hypothesize that the ordering process may be due to a restriction of the more energetic $C_2$ and $C_4$ reorientations (or modes) at a critical temperature $T_C$, of which immediately below, the $\mu_p$ of each $NH_4^+$ can only point along 8 directions instead of 14. This causes a problem because adjacent $NH_4^+$ ions have a high probability that the directions of their $\mu_p$ are the same which causes the two ions to become mutually resonant with each other. This situation destabilizes the lattice and the only means to avoid structural collapse is for the $\mu_p$ to establish a spatial modulation to distribute the resonant forces more evenly. A simpler way to picture this is to treat the reorienting $NH_4^+$ as tiny bar magnets situated at the center of a pseudo-cube. When the sample is cooled to $T_C$, the directions of the bar magnets are only allowed to point along the corners of the pseudo-cube. From such, adjacent $NH_4^+$ exhibit forces on each other which induces strain on the lattice. Globally, the directions of the bar magnets must exhibit a type of modulation (become ordered) otherwise unevenly distributed forces will rupture the lattice periodicity. Above $T_C$, the directions of the bar magnets can also point along the faces of the pseudo-cube ($C_2$ and $C_4$ allowed) and fluctuate enough so that the bar magnets do not need to become ordered. The *ABAB* pattern along the *ab*-axis in the $\gamma$ phase is therefore a byproduct of $NH_4^+$ arranging into a configuration that cancels out the resonant magnetic forces. An effect of the *ABAB* pattern is a slight alternating offset of the halide anions along the ±*c* axis [9]. Therefore, equivalent electric dipole moment ***p*** elements reside between every $NH_4^+$ site and the $\gamma$ phase may be regarded as antiferroelectric along the *c*-axis. Figure 1d shows a spiral wavevector 8 unit-cells long (green line) representing a possible spatial modulation of the directions of $\mu_p$ (grey dashed line) near the $I^-$ positions. Ordered magnetic spirals can induce a macroscopic polarization ***P*** because an inversion of the coordinate of the propagation vector will also invert the magnetic moments, therefore breaking both time-reversal and spatial-inversion symmetries [10,11]. However, in our molecular analogue, adjacent rows will have magnetic spirals in the opposite direction because of the *ABAB* pattern so the sum of all ***p*** remains zero in superstructures of 2 x 4 unit-cells in the *ab*-planes. Suppose now that an $NH_4^+$ were to be replaced by a $K^+$ ion, then a local absence of $\mu_p$ will cause ***p*** to decrease at the local site and render ***P*** $\neq 0$. If the replacement of $NH_4^+$ by $K^+$ is a multiple of 1:8, then the remnant polarization of neighboring superstructures can end up statistically cancelling each other out. However, if the $K^+$ to $NH_4^+$ ratio is not a multiple of 1:8, for instance, 1:9 or 1:7, then there will be a subset of superstructures with no counterparts to cancel their remnant polarizations. Hence, a generalization may be obtained in that if the $NH_4^+$ to $K^+$ ratio is incommensurate to the size of the superstructure, then the ordered phase of $(NH_4)_{1-x}K_xI$ should be polar.

The phase diagram of $(NH_4)_{1-x}K_xI$ has been studied by several research groups [12,13] (Fig. 2). Pure $NH_4I$ is cubic of the NaCl type ($\alpha$ phase) at room temperature and phase transitions to cubic of the CsCl type ($\beta$ phase) upon cooling at $T_{\alpha\text{-}\beta}$ = 256 K and to slightly tetragonal ($\gamma$ phase) at $T_{\beta\text{-}\gamma}$ = 232 K [14]. With replacement of small concentrations of $K^+$ up to *x* = 0.18, only an NaCl and a CsCl phase can be stabilized at high and low temperatures, respectively. With *x* > 0.18, the CsCl phase is quenched as the NaCl phase was reported to remain stable even down to 10 K [12]. With even higher concentrations of *x* > 0.38 the system becomes an orientational glass [15-17] where the correlation length is reduced to near 10 unit-cells (for the deuterated analogue) characterized by short-range antiferroelectric correlations [12]. From such, according to our hypothesis, samples with $K^+$ concentrations of 0 < *x* < 0.18 may exhibit ferroelectricity or pyroelectricity at the very least. To our knowledge, no researchers have taken into account the possible generation of a macroscopic polarization via an incommensurate correlation of reorienting aspherical and near-stationary spherical ions with respect to the size of superstructures based on proton orbitals. This explains why only the dielectric constant of $(NH_4)_{1-x}K_xI$ crystals have been measured before while no reports on whether the compounds are pyroelectric and magnetic have been made. In this work, we show through pyroelectric current and dielectric constant measurements that $(NH_4)_{0.95}K_{0.05}I$ crystals are ferroelectric while $(NH_4)_{0.87}K_{0.13}I$ and $(NH_4)_{0.83}K_{0.17}I$ crystals are pyroelectric. Pronounced discontinuities in the magnetic susceptibility coinciding to the peaks in the electric polarization and dielectric constant indicate that the order-disorder phase transition is driven by proton orbital-orbital interactions. With the myriad of possibilities of partially replacing compounds with the $NH_4^+$ ion by spherical ions (such as $Na^+$, $K^+$, $Rb^+$ or $Cs^+$), or the other way around, researchers can employ the same approach to further enhance the functionalities of advanced materials and improve structural integrity of existing $NH_4^+$–based molecular framework materials via reduction of internal magnetic resonance.

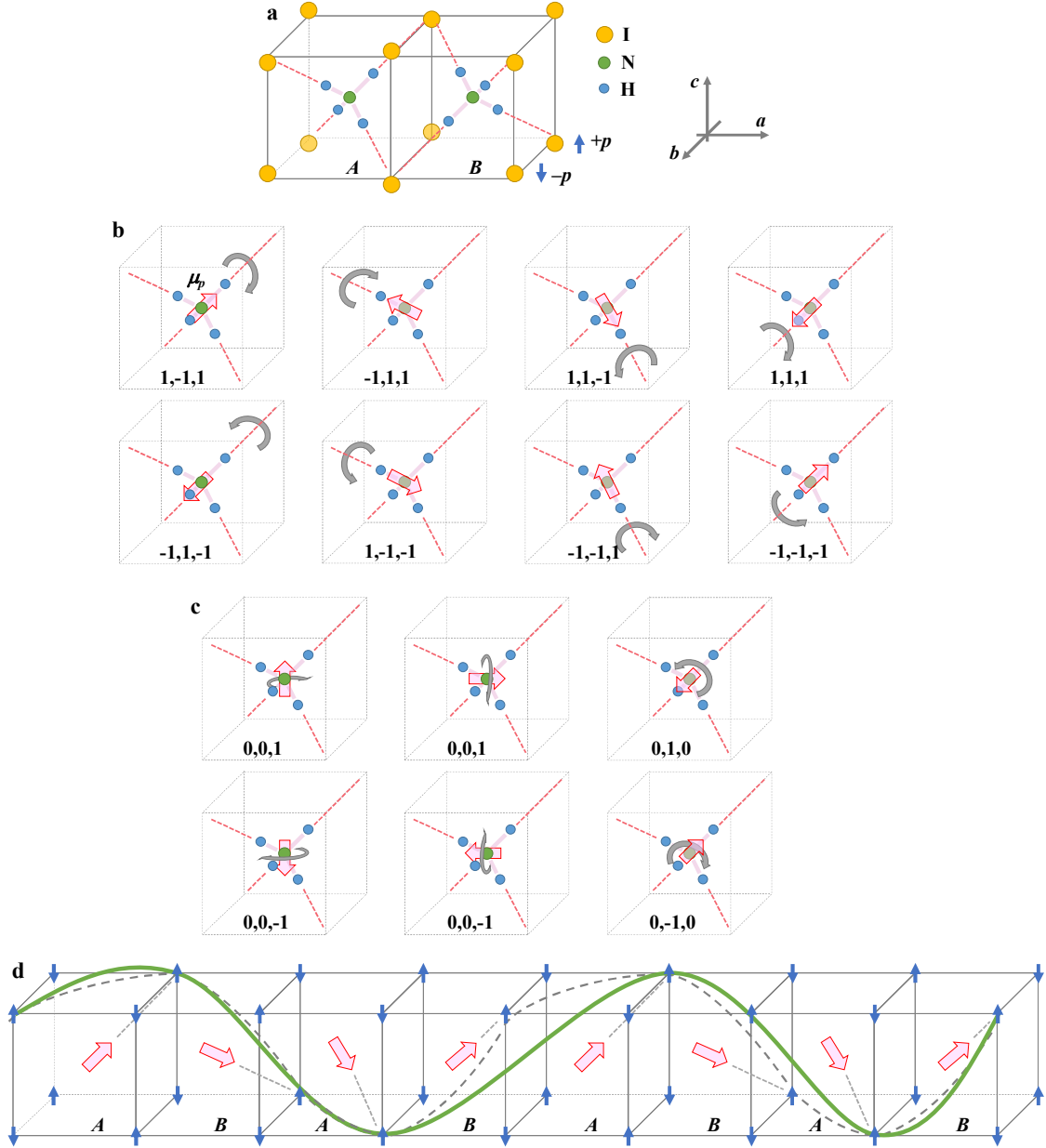

**Fig. 1.** Spatial and temporal structures of the ordered $\gamma$ phase of $NH_4I$. (a) All $NH_4^+$ take one of two equilibrium spatial configurations $A$ and $B$. The $I^-$ ions are offset by 0.0012 Å (Ref. [9]) along the $c$-axis so there also exist local electric dipole moments $\pm p$. Pink dashed lines represent hydrogen bonds. (b) The 8 possible types of $C_3$ reorientations when each $NH_4^+$ rotate by 120° about an N–H bond. The associated orbital magnetic moment $\mu_p$ traced out by the protons point along the diagonals of the pseudo-cube. (c)

The 6 possible types of $C_2$ and $C_4$ reorientations when entire $NH_4^+$ rotate about the *a*-, *b*- or *c*-axes; their respective $\mu_P$ point along the faces of the pseudo-cube. d) Only the $C_3$ reorientations are believed to continue in the $\gamma$ phase so the $NH_4^+$ arrange into an *ABAB* pattern along the *a*- and *b*-axes to evenly distribute the internal orbital resonances of the proton orbitals. Dashed line represents the spatial modulation of the strength of $\mu_P$ at the $I^-$ positions represented by *p* which can be roughly fitted by a spiral wave vector (green solid line).

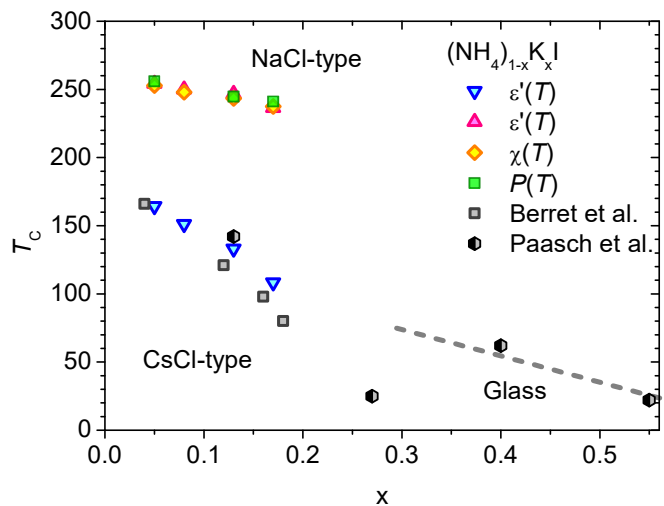

**Fig. 2.** Phase diagram of $(NH_4)_{1-x}K_xI$ according to Berret *et al.* [12] and Paasch *et al.* [13] along with data points obtained from this work (colored symbols) via electric polarization $P(T)$, dielectric constant $\varepsilon'(T)$ and magnetic susceptibility $\chi(T)$. Dashed line represents the glass transition temperature.

## 2. Materials and methods

Ammonium iodide $NH_4I$ (99.99% in purity) and potassium iodide KI (99.99% in purity) from Aladdin-e, Inc. (Shanghai) were mixed in different molar ratios in deionized water; the solution was slowly evaporated to obtain crystals of $(NH_4)_{0.95}K_{0.05}I$, $(NH_4)_{0.87}K_{0.13}I$ and $(NH_4)_{0.83}K_{0.17}I$. The method to determine the $NH_4^+$ and $K^+$ concentrations of the samples is available in the Supplementary Material file. The pyroelectric current was measured by a Keithley 6517B electrometer. Two electrodes were first applied along the *c*-axis of the samples resembling a parallel plate capacitor. A poling electric field was applied at room temperature and maintained down to 100-160 K. The two electrodes were then shorted for 30 minutes and its pyroelectric current was measured during warming. The complex dielectric constant was obtained from the measured capacitance and loss tangent via the same pair of electrodes but with an Agilent E4980A LCR meter. The dimensions of the samples for the electric measurements were near 2 x 4 x 0.5 $mm^3$. The magnetic susceptibility was measured by the Vibrating Sample Magnetometer option of a Physical Properties Measurement System made by Quantum Design, Inc. (San Diego). The size of samples for the magnetic measurements were ~2 x 4 x 1.5 $mm^3$. The cooling and warming rates in all experiments were between 1 and 2 K/min.

## 3. Results and discussion
### 3.1 Electric polarization measurements

Figure 3a shows the electric polarization $P(T)$ of $(NH_4)_{1-x}K_xI$ for when $x = 0.05$ obtained by numerical integration of the measured pyroelectric current $I(T)$ (inset of Fig. 3a). All samples (including the $x = 0.13$ and 0.17 ones) were measured during warming after application of a poling field of $+E = +0.5$ kV/cm during cooling. Then, immediately afterwards the same process was repeated but with a field of negative bias $-E = -0.5$ kV/cm to check whether the polarity is reversible. On both poling occasions for when $x = 0.05$ a maximum polarization of $\pm 0.26$ µC/cm$^2$ was observed. As a comparison, the polarization of Rochelle salt at 278 K is 0.25 µC/cm$^2$ [18]. Returning to the $x = 0.05$ sample, two peaks in $I(T)$ were observed to occur on average at 228 K and 256 K. At such low level of K$^+$ concentration, the two transition temperatures appear to represent the $\gamma$-$\beta$ and $\beta$-$\alpha$ phase transitions which classify the $\beta$ and $\gamma$ phases in $(NH_4)_{0.95}K_{0.05}I$ as ferroelectric. For this reason, only the point 0.05, 256 K is plotted in Fig. 2 to represent the onset of spontaneous polarization.

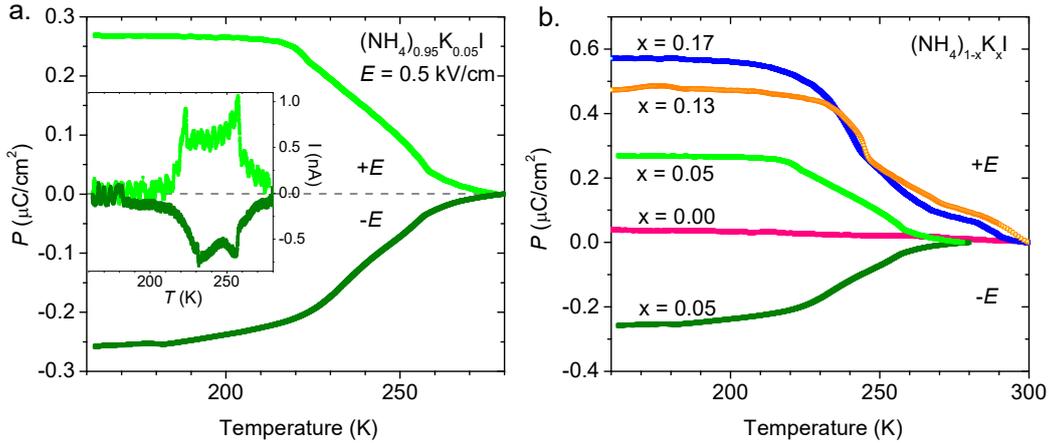

**Fig. 3.** Electric polarization $P(T)$ of $(NH_4)_{1-x}K_xI$ when $x = 0$, 0.05, 0.13 and 0.17. (a) $P(T)$ of the $x = 0.05$ crystal under poling fields of $+E = +0.5$ and $-E = -0.5$ kV/cm. Inset shows the pyroelectric current from which $P(T)$ was integrated from. (b) $P(T)$ for the cases when $x = 0$, 0.13, and 0.17 under $+E$ and 0.05 under $\pm E$.

Figure 3b shows in addition the $P(T)$ curves of the crystals $x = 0$, 0.13 and 0.17 under positive poling. The peaks in $I(T)$ for when $x = 0.13$ and 0.17 occurred at 244.4 K and 241.2 K, respectively, which are taken as the CsCl to NaCl transition temperatures in Fig. 2. For these two samples, only the $P(T)$ curves under $+E$ are shown because the polarity was not reversible under $-E$ so these two systems can only be classified as being pyroelectric. The pure NH$_4$I crystal was not pyroelectric as expected since its $\alpha$, $\beta$ and $\gamma$ phases have space groups $Fm3m$, $Pm3m$ and $P4/nmm$, respectively, all of which are non-polar.

### 3.2 Complex dielectric constant measurements

Figures 4a to 4f show the real $\varepsilon'(T)$ and imaginary $\varepsilon''(T)$ parts of the dielectric constant measured under 1 kHz for when $x = 0.05$, 0.13 and 0.17 during cooling and warming. The most pronounced features are the different peaks that only appear during warming; all maxima are in good agreement with the peaks observed in $I(T)$. In the case of the $(NH_4)_{0.95}K_{0.05}I$ sample, only the smaller peak at 254.2 K observed in $\varepsilon'(T)$ is plotted in the phase diagram (Fig. 2) as this seems to be the critical temperature representing the order-disorder transition in agreement with $I(T)$ and the magnetic susceptibility data presented below. Another set of discontinuities in the form of a step-down anomaly in $\varepsilon'(T)$ were observed during cooling only (insets of Figs. 4a, 4b and 4c). These anomalies exhibited a systematic decrease in the critical temperature with respect to $x$ which coincides nearly exactly to the disorder-to-order phase boundary line in the phase diagram of $(NH_4)_{1-x}K_xI$ by Berret *et al*. [12]. The difference between the cooling and warming curves of $\varepsilon'(T)$ and $\varepsilon''(T)$ may be due to the rather large range of hysteretic behavior characteristic of the ammonium halides as the CsCl and NaCl phases share metastable regions of nearly 100 K [19,20]. Moreover, some phases can coexist for a large temperature range [21] which also explains the non-zero polarization values above the phase transitions. The Supplementary Material file shows the measured $\varepsilon'(T)$ and $\varepsilon''(T)$ of an $(NH_4)_{0.95}K_{0.05}I$ sample in powdered form (Figs. S1a and S1b) which showed similar results to those in Figs. 4a and 4d. This allowed us to rule out the possibility of interfacial and surface effects as well as sample degradation to be the reason why the cooling and warming curves of $\varepsilon'(T)$ and $\varepsilon''(T)$ are different. Lastly, it is worth noting that the magnitudes of $P(T)$ and $\varepsilon'(T)$ do not necessarily have to exhibit a systematic change because the polarization of neighboring superstructures cancel each other out at specific concentrations of $x = n / \xi$, where $n = 0, 1, 2 \ldots$ and $\xi$ is the size of the superstructures.

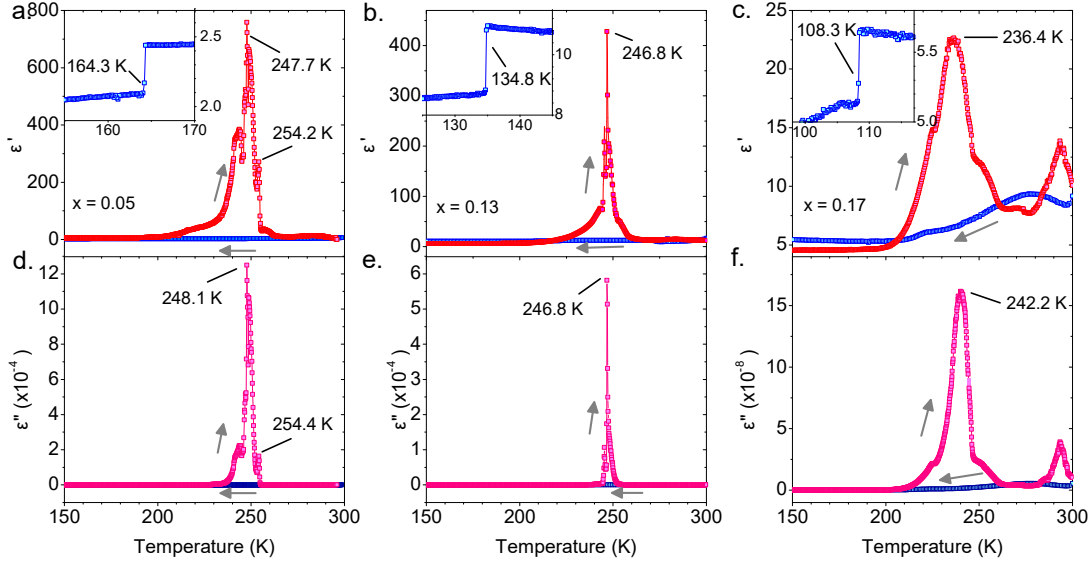

**Fig. 4.** (a)-(c) Real ε'(*T*) and (d)-(e) imaginary ε''(*T*) parts of the dielectric constant at 1 kHz for the cases when $x$ = 0.05, 0.13 and 0.17. Insets in (a)-(c) show the cooling curves of ε'(*T*) near the NaCl to CsCl transition.

### 3.3 Magnetic susceptibility measurements

Figure 5 shows the measured molar magnetic susceptibility $\chi(T)$ for single crystals of pure $NH_4I$ and the three sample concentrations of $x$ = 0.05, 0.13 and 0.17 during cooling and warming at rates of 1 K/min under an external magnetic field of $H$ = 10 kOe. No pronounced discontinuities were observed in $\chi(T)$ in the mixed samples when cooled across the NaCl to CsCl phase transition; the process appeared to occur gradually. However, a sharp step-down discontinuity was always observed when the samples were warmed back across the CsCl to NaCl transition. This was especially evident for the pure $NH_4I$ sample where $\chi(T)$ increased in magnitude by ~9%. The volume change across two different types of cubic and pseudo-cubic structures is usually less than 0.1% so the step-anomalies can only be attributed to stem from the order-disorder process of the proton orbitals. The size and location of the critical points of these steps progressively decreased with $K^+$ concentration. The transition temperatures of these critical points are also plotted in Fig. 2 and are in good agreement with the *P*(*T*) and ε'(*T*) results.

A noticeably odd feature in the $\chi(T)$ curves is that the cooling curves are always smaller in magnitude than the warming curves. As the samples are cooled, microcracks develop because not enough time is allowed for the resonant forces to re-adjust. The cracks are visible with the naked eye from the bleaching of the crystals after each cooling cycle. From such, during warming the original crystal essentially becomes a collection of slightly misaligned polycrystalline samples so if there exists magnetic anisotropy such as a different χ along the 1,1,1 and 1,0,0 directions, then the cooling and warming curves are bound to be slightly different. These results indicate that the $NH_4^+$–$NH_4^+$ interactions in such types of "plastic crystals" may also possess a magnetic nature rather than only electric dipolar, quadrupolar, octupolar and/or van der Waals interactions [21-24].

Another possible scenario regarding the difference between the cooling and warming curves in not only $\chi(T)$ but also in $\varepsilon'(T)$ and $\varepsilon''(T)$ is that the disorder to order process should require more time to complete than the order to disorder process since in the former case neighboring $NH_4^+$ need to negotiate and figure out the direction of the global modulation vector. In this process, various nucleation sites and therefore domains of the ordered phase with different modulation vector directions should form. It may well be that if the sample was cooled extremely slow, then a sharp discontinuity in $\chi(T)$ as well as peak anomalies in $\varepsilon'(T)$ and $\varepsilon''(T)$ should also be observed, similar to their counterparts during warming. In contrast, the order to disorder process of the $NH_4^+$ is more abrupt because there is no sudden increase in energy degeneracies that need to be 'lifted' so the transitions during warming were resolvable under warming rates of 1 K/min.

Lastly, the observed features appear to not originate from paramagnetic impurities because a) small traces of impurities would render $\chi(T)$ to become positive; b) the warming and cooling curves would not exhibit mismatches; and c) a paramagnetic tail proportional to $T^{-1}$ would dominate at low temperatures.

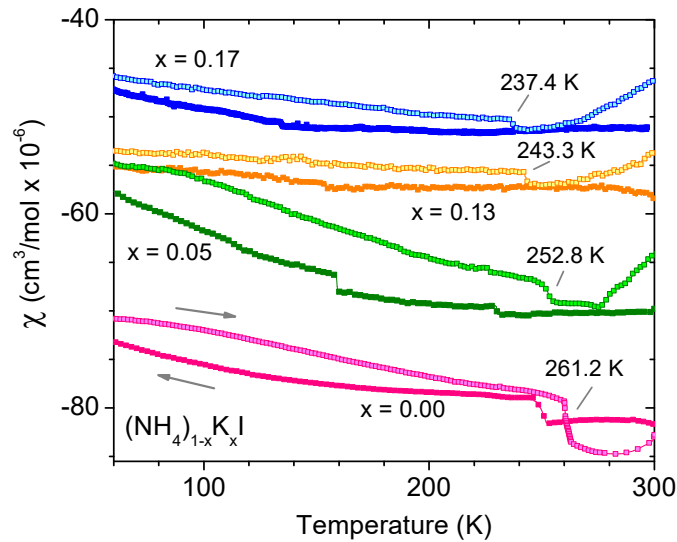

**Fig. 5.** Molar magnetic susceptibility of $(NH_4)_{1-x}K_xI$ when $x = 0$, 0.05, 0.13 and 0.17 during cooling and warming plotted with darker and lighter colors, respectively, under an external magnetic field of $H = 10$ kOe.

One of the ways ferroelectricity is induced in crystals is when a lattice mode

'freezes' below a critical temperature: for instance, when a particular atom such as nitrogen is no longer energetically favorable to jump back and forth between two potential wells. In the current case, the freezing of the $C_4$ modes cause the remaining $C_3$ modes to build up internal magnetic resonance in the lattice and it is the re-adjusting of the coupled electric dipole moments (unable to fully cancel out) that generates a polar phase. This quantitatively explains why the polarization is nearly impossible to switch below the ferroelectric Curie temperature $T_C$ because a bias electric field is unable to change all of the $A$ configurations to $B$ or vice versa without the $C_4$ and $C_2$ modes active. This perspective can be applied to explain why several other $NH_4^+$ compounds that are claimed to be ferroelectric are through pyroelectric current measurements or other methods while $P$-$E$ hysteresis loops are mostly absent [25-29]. The mechanism behind the polar phases of these hydrogen-based materials seems to be a molecular orbital-lattice type of coupling instead of the conventional electron spin-orbit coupling [10] and a full theoretical treatment is certainly needed. Nevertheless, there exists countless of other $NH_4^+$ compounds that exhibit order-disorder phase transitions at higher temperatures [30–33]. They usually exhibit λ-peaks in the specific heat at the order-disorder transition and are good candidates to replace a small amount of their $NH_4^+$ ions by spherical ions to potentially create entire families of exotic ferroelectrics.

## 4. Conclusions

To conclude, we find that the CsCl phase of $NH_4I$ becomes ferroelectric when 5% of its ammonium cations are replaced by $K^+$ cations. With replacement of up to 17% of $K^+$, $(NH_4)_{1-x}K_xI$ continues to be pyroelectric and has the potential of becoming ferroelectric. At this point, we can only speculate that the mechanism underlying the onset of spontaneous polarization appears to be due to an incommensurate ratio of $K^+$ and $NH_4^+$ ions with respect to the size of the proton-orbital superstructure resulting in the associated molecular magnetic moments, which are coupled to the off-centered $I^-$, to not completely cancel out. These assumptions may be verified by studying the structure of $(NH_4)_{1-x}K_xI$ thin films. A similar approach can be applied to other types of molecular frameworks and potentially render them multiferroic. By the same means, our methodology should increase the small pool of highly desirable organic ferroelectrics [34].

**Acknowledgements:**

This work was supported by the Science, Technology and Innovation Commission of Shenzhen Municipality, a General Project grant of the Shenzhen Universities Sustained Support Program #GXWD20201230155427003-20200822225417001.